\documentclass[prb,twocolumn,a4paper,superscriptaddress,showpacs,floatfix]{revtex4-1}

\usepackage{dcolumn,amsmath,xspace}
\PassOptionsToPackage{caption=false}{subfig}
\usepackage{subfig}
\usepackage{graphicx}
\usepackage{multirow}
\usepackage{dcolumn}
\usepackage{amssymb}
\usepackage{latexsym}

\usepackage[utf8]{inputenc}
\usepackage{lineno}

\usepackage{epstopdf}

\newcommand{\ums}{\ensuremath{\mu\mathrm{m}^2}\xspace}
\newcommand{\us}{\ensuremath{\mu\mathrm{s}}\xspace}
\newcommand{\sub}[1]{\ensuremath{_\mathrm{#1}}}

\begin{document}

\title{Time-resolved PhotoEmission Spectroscopy on a Metal/Ferroelectric Heterostructure}

\author{J. E. Rault}
\altaffiliation[Now at: ]{Synchrotron-SOLEIL, BP 48, Saint-Aubin, F91192 Gif sur Yvette CEDEX, France}
\affiliation{CEA, DSM/IRAMIS/SPCSI, F-91191 Gif-sur-Yvette Cedex, France}
\author{G. Agnus}
\affiliation{Institut d'Electronique Fondamentale, Univ Paris-Sud, CNRS UMR 8622, 91405 Orsay Cedex, France}
\author{T. Maroutian}
\affiliation{Institut d'Electronique Fondamentale, Univ Paris-Sud, CNRS UMR 8622, 91405 Orsay Cedex, France}
\author{V. Pillard}
\affiliation{Institut d'Electronique Fondamentale, Univ Paris-Sud, CNRS UMR 8622, 91405 Orsay Cedex, France}
\author{Ph. Lecoeur}
\affiliation{Institut d'Electronique Fondamentale, Univ Paris-Sud, CNRS UMR 8622, 91405 Orsay Cedex, France}
\author{G. Niu}
\altaffiliation[Now at: ]{IHP, Im Technologiepark 25, D-15236 Frankfurt (Oder), Germany}
\affiliation{Université de Lyon, Ecole Centrale de Lyon, Institut des Nanotechnologies de Lyon, F-69134 Ecully cedex, France}
\author{B. Vilquin}
\affiliation{Université de Lyon, Ecole Centrale de Lyon, Institut des Nanotechnologies de Lyon, F-69134 Ecully cedex, France}
\author{M. G. Silly}
\affiliation{Synchrotron-SOLEIL, BP 48, Saint-Aubin, F91192 Gif sur Yvette CEDEX, France}
\author{A. Bendounan}
\affiliation{Synchrotron-SOLEIL, BP 48, Saint-Aubin, F91192 Gif sur Yvette CEDEX, France}
\author{F. Sirotti}
\affiliation{Synchrotron-SOLEIL, BP 48, Saint-Aubin, F91192 Gif sur Yvette CEDEX, France}
\author{N. Barrett}
\email[Correspondence should be addressed to ]{nick.barrett@cea.fr}
\affiliation{CEA, DSM/IRAMIS/SPCSI, F-91191 Gif-sur-Yvette Cedex, France}

\begin{abstract}
In thin film ferroelectric capacitor the chemical and electronic structure of the electrode/FE interface can play a crucial role in determining the kinetics of polarization switching. We investigate the electronic structure of a Pt/BaTiO\sub{3}/SrTiO\sub{3}:Nb capacitor using time-resolved photoemission spectroscopy. The chemical, electronic and depth sensitivity of core level photoemission is used to probe the transient response of different parts of the upper electrode/ferroelectric interface to voltage pulse induced polarization reversal. The linear response of the electronic structure agrees quantitatively with a simple RC circuit model. The non-linear response due to the polarization switch is demonstrated by the time-resolved response of the characteristic core levels of the electrode and the ferroelectric. Adjustment of the RC circuit model allows a first estimation of the Pt/BTO interface capacitance. The experiment shows the interface capacitance is at least 100 times higher than the bulk capacitance of the BTO film, in qualitative agreement with theoretical predictions from the literature.
\end{abstract}

\vspace*{4ex}

\maketitle
\newpage


\section{Introduction}

Epitaxially strained oxide thin films make ferroelectric (FE) based heterostructures of real interest for new electronic devices~\cite{ramesh_multiferroics:_2007, schlom_thin_2008}. High read and write speeds are of prime importance for implementing ferroelectric based electronics. The dynamic process of polarization reversal is crucial to understand the application limits in terms of maximum frequency and long-term fatigue~\cite{balke_direct_2010}. It is therefore essential to understand the polarization switching kinetics in thin film ferroelectric capacitor structures. 

Polarization reversal in the prototypical BaTiO$_\mathrm{3}$ ferroelectric singles crystals was first studied by W.J.~Merz~\cite{merz_domain_1954}. The switching dynamics have been described by classical nucleation theory of Kolmogorov, Avrami and Ishibashi in which the polarization is reversed first by domain nucleation then domain wall motion~\cite{ishibashi_note_1971, avrami_kinetics_1939, kolmogoroff_zur_1937}. In smaller capacitive stuctures, nucleation limited switching have been found to provide a more accurate description~\cite{tagantsev_non-kolmogorov-avrami_2002, gruverman_piezoresponse_2008, zhukov_dynamics_2010}. Following the model of Merz, the switching time depends exponentially on the activation field~\cite{merz_domain_1954}. Gruverman \textit{et al.} have shown that there are high and low field regimes for polarization reversal. Small capacitive structures under a high field switch more rapidly whereas larger area capacitors switch more quickly under low fields because the kinetics are nucleation limited~\cite{gruverman_piezoresponse_2008}.

Most characterization of FE switching has been done using electrical methods but accurate switching measurements are often limited by the RC circuit load on the device and parasitic resistances. Despite this problem, it has been shown that the activation field can be determined independently of the RC circuit characteristics, the effect of the latter being a time dilation of the FE switch~\cite{song_activation_1998}.
 
The properties of the electrode/FE interface may also determine the switching kinetics. For example, it has been suggested that defect related local fields determine the switching time in BiFeO$_\mathrm{3}$ capacitors~\cite{kim_polarity-dependent_2011}. At the electrode/FE interface, chemistry, strain and electronic ordering combine in a complex way to screen the interface polarization charge of the ferroelectric layer. An accurate description of the dynamical electronic structure may help to better understand the switching process and the phenomena underlying the results of electrical measurements. However, experimental data on the electronic properties of such systems are scarce, principally due to the intrinsic difficulty of measuring a buried interface.

The use of complementary \textit{in-operando}, \textit{i.e.} submitted to a realistic voltage excitation, structural characterization techniques is therefore desirable. Ideally, these techniques should be independent of the electrical circuit required to switch the polarization. For instance, Gorfman \textit{et al.} used time-resolved x-ray diffraction to determine the lattice dynamics and atomic positions in piezoelectrics~\cite{gorfman_time-resolved_2010}.

X-ray photoemission spectroscopy (XPS) is a powerful, direct tool for characterizing the chemical and electronic interface structure. It has been used with \textit{in-situ} bias to demonstrate the migration of oxygen vacancies in Pt/HfO\sub{2}~\cite{nagata_oxygen_2010} or to conduct impedance-like measurements on Rb deposited on silicon~\cite{suzer_impedance-type_2009}. Chen and Klein used \textit{in-situ} photoemission spectroscopy on a BTO monocristal to deduce the barriers properties as a function of ferroelectric polarization~\cite{chen_polarization_2012}. Our group studied a Pt/BaTiO$_\mathrm{3}$/Nb:SrTiO$_\mathrm{3}$ (Pt/BTO/NSTO) heterostructure with bias application and found polarization dependent barriers and a non-trivial behavior of the core-levels as a function of applied bias~\cite{rault_interface_2013}. All of these studies show the potential of photoemission experiments for \textit{in-operando} work.

An exciting route for a fundamental understanding of the interface electronic structure of FE devices would therefore be time-resolved photoelectron spectroscopy. State-of-the-art delay-line detectors allow detection of photoemitted electrons with a time-resolution down to 5 to 10~ns~\cite{bergeard_time-resolved_2011}. Although switching times can be of the order of a nanosecond~\cite{larsen_nanosecond_1991}, in many cases it may be sufficient to resolve the transient response of the electronic structure to the applied voltage pulse. The aim is to detail the transient response of the electronic structure to the switching pulse and the role of the interface and bulk film capacitance in defining ultimate device speed.

We investigate the electronic structure of an electrode/FE interface using time-resolved photoemission spectroscopy. The static properties of the ferroelectric capacitance have been studied by photoemission in Ref.~\onlinecite{rault_interface_2013}. Bias pulses on a Pt/BTO/NSTO capacitor probe the time-resolved chemical and electronic structure of the Pt/BTO interface. An equivalent circuit model is used predicting a behavior which compares well with the photoemission results on the real system. This shows the potential of time-resolved photoemission spectroscopy to follow the chemical/electronic changes in working model FE microelectronic devices. 

\section{Experiment}

\begin{figure}[h]
  \centering
   	\subfloat{\label{fig:dispo_scheme}\includegraphics[scale=0.46]{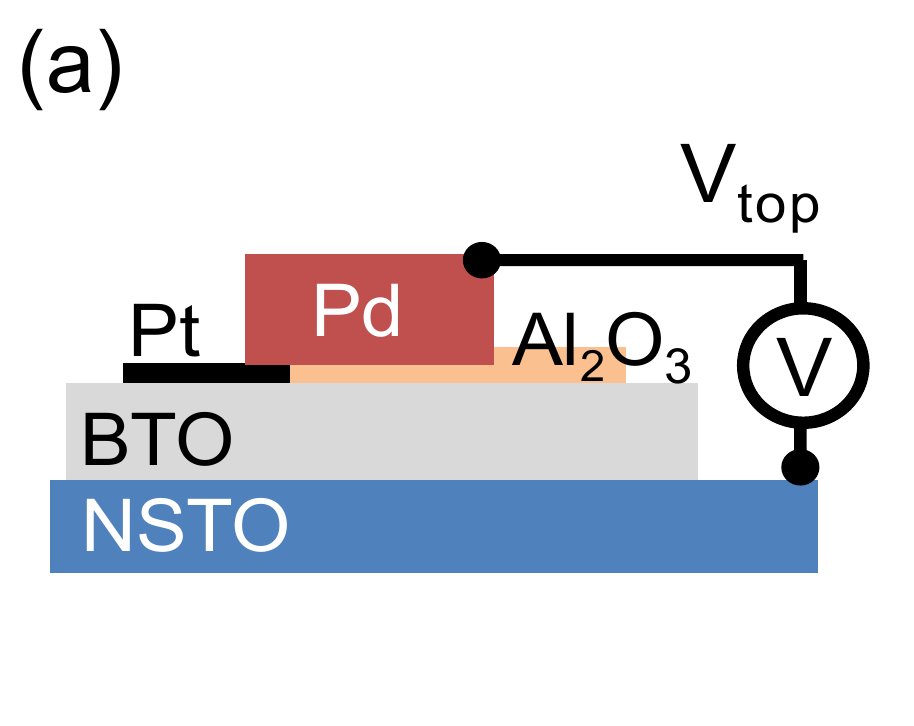}}\hfill
	\subfloat{\label{fig:map_Pt}\includegraphics[scale=0.46]{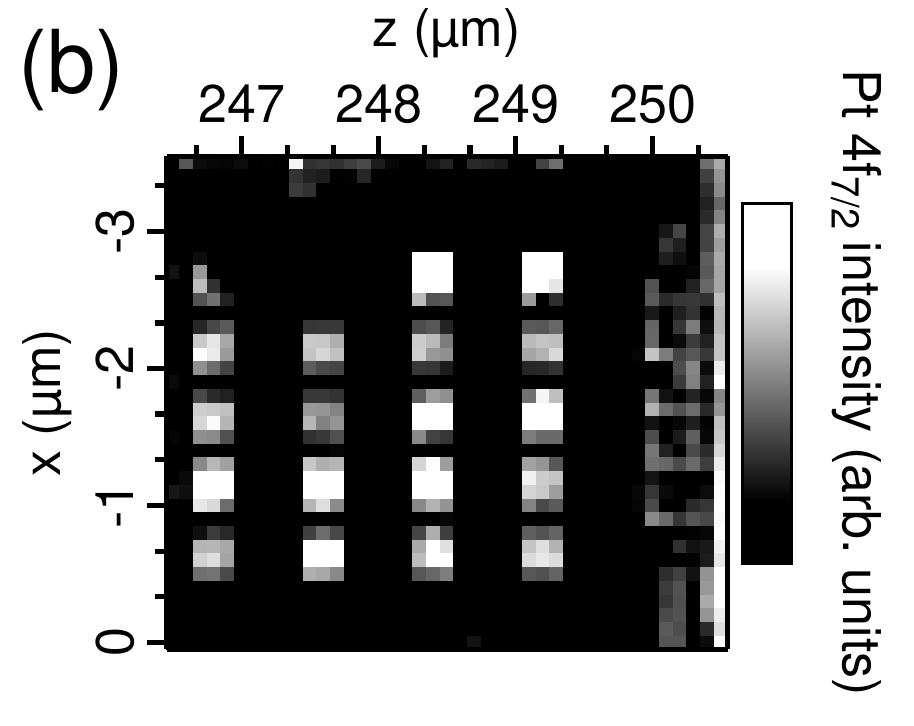}}\\
  \caption{(a) Schematic of the capacitor; (b) Pt~$4f$ intensity map for the Pt/BTO/NSTO sample showing 20 identical Pt/BTO/NSTO capacitors ($300 \times 300$~\ums) on the $5 \times 5$~mm$^2$ surface, allowing location of the wired capacitor.}
  \label{fig:dispositif}
\end{figure}

The sample is a Pt (2.8~nm)/BTO (64~nm)/NSTO heterostructure grown by Molecular Beam Epitaxy. The growth conditions and heterostructure properties are described in Ref.~\onlinecite{rault_interface_2013}. $300 \times 300$~\ums thick Pt electrodes were patterned by ionic beam etching. Thicker (300 nm) Palladium (Pd) pads overlapping part of the Pt electrodes have been deposited by evaporation to enable wire-bonding of the top electrodes to the sample holder. A highly insulating layer of Al$_\mathrm{2}$O$_\mathrm{3}$ was deposited by evaporation onto bare BTO to suppress interference of the Pd pads with the capacitance. The device architecture is fully described in Ref.~\onlinecite{rault_interface_2013} and is shown in Fig.~\ref{fig:dispo_scheme}. When the polarization is pointing from the bottom to top electrode (P+) the leakage current is limited by Schottky emission at the upper, Pt/BTO interface whereas in the P- state the current flow is determined by the quasi-ohmic BTO/NSTO interface. The device was then introduced in ultrahigh vacuum (10$^\mathrm{-8}$~Pa) in the XPS set-up of the TEMPO Beamline at the SOLEIL synchrotron radiation source~\cite{polack_tempo:_2010}. The $100 \times 100$~\ums beam could be directed onto a single top electrode located by a map of the whole sample using the Pt absorption edge (see Fig.~\ref{fig:map_Pt}). A photon energy of 1100~eV was used to optimize the signal from the BTO close to the interface (estimated probing depth of $\approx$3-4~nm). The overall energy resolution was 220~meV. The sample-holder allows \textit{in-situ} bias application and electrical measurements via high quality coax wires to limit parasitic behavior due to the electrical environment.

The coercive voltage V$_\mathrm{c-}$ needed to switch from P$^{+}$ to P$^{-}$ is 0.80~V and V$_\mathrm{c+}$, to switch from P$^{-}$ to P$^{+}$, is 0.40~V. The loop shows a 0.6~V offset with a P$^{+}$ remnant state at zero voltage. In order to ensure fully switched polarization states we apply voltages beyond the coercive values,  V$_\mathrm{top} = 0.35$~V ($0.85$~V) for the P$^{+}$ (P$^{-}$ state). To investigate the time-resolved properties, we use a train of voltages pulses. One is a non-switching pulse, the other is a switching pulse. This design allows to discriminate phenomena due to ferroelectric polarization (non-linear ferroelectric behavior, switching pulse) from those due to the whole device (linear dielectric behavior, non-switching pulse)~\cite{traynor_capacitor_1997}.

\begin{figure}[!ht]
  \centering
  	\includegraphics[scale=0.50]{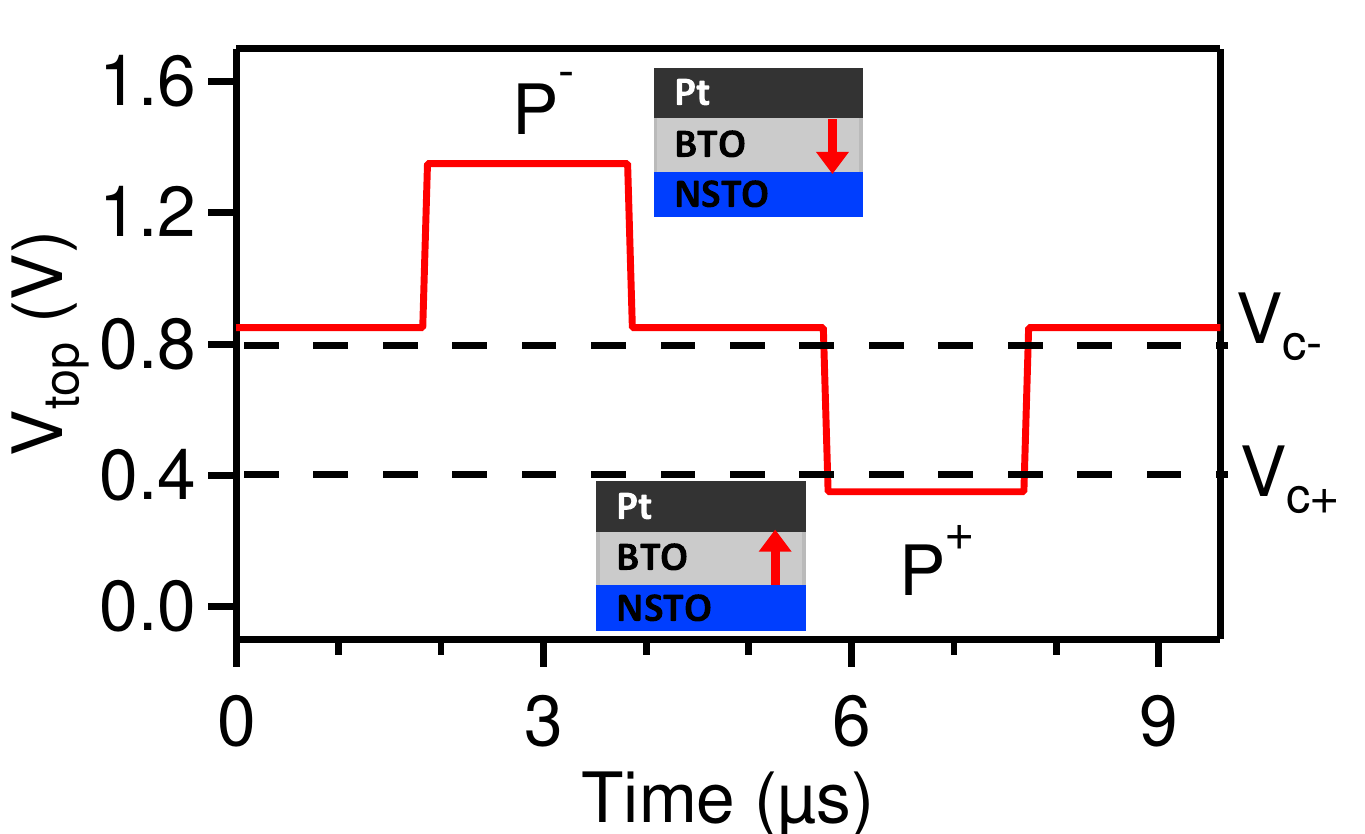}
  \caption{V$_\mathrm{top}$ pulse train as a function of time. The first pulse is non-switching P$^{-}$ to P$^{-}$, the second is switching P$^{-}$ to P$^{+}$ then back to P$^{-}$. The BTO film is in the P$^{+}$ state between $t = 5.8$~\us and $t = 7.8$~\us. The coercive voltages V$_\mathrm{c+}$ and V$_\mathrm{c-}$ are indicated by dotted lines.}
  \label{fig:source_tr}
\end{figure}

At $t = 0$, we set V$_\mathrm{top} = 0.85$~V, the idle state, which corresponds to the P$^{-}$ polarization. At $t = 1.9$~\us, an upward pulse V$_\mathrm{top} = 1.35$~V is applied with a rise (fall) time of 2.5~ns and a width of 2~\us, then V$_\mathrm{top}$ goes back to the idle state. Thus, during this non-switching pulse, the system stays in the P$^{-}$ state. At $t = 5.8$~\us, a downward pulse V$_\mathrm{top} = 0.35$~V is applied and the system switches from P$^{-}$ to P$^{+}$. At $t = 7.8$~\us, V$_\mathrm{top}$ returns to the idle state and the system switches from P$^{+}$ to P$^{-}$ (see Figure~\ref{fig:source_tr}).
Snapshot spectra of the photoemitted electrons are acquired with a time step of 45~ns over 9~\us. The pulse train is repeated a thousand times in order to acquire a sufficient photoemission signal/noise ratio.

\section{Results}

\subsection{Photoemission Data}

\begin{figure}[!ht]
  \centering
   	\subfloat{\label{fig:tr_Ba}\includegraphics[scale=0.5]{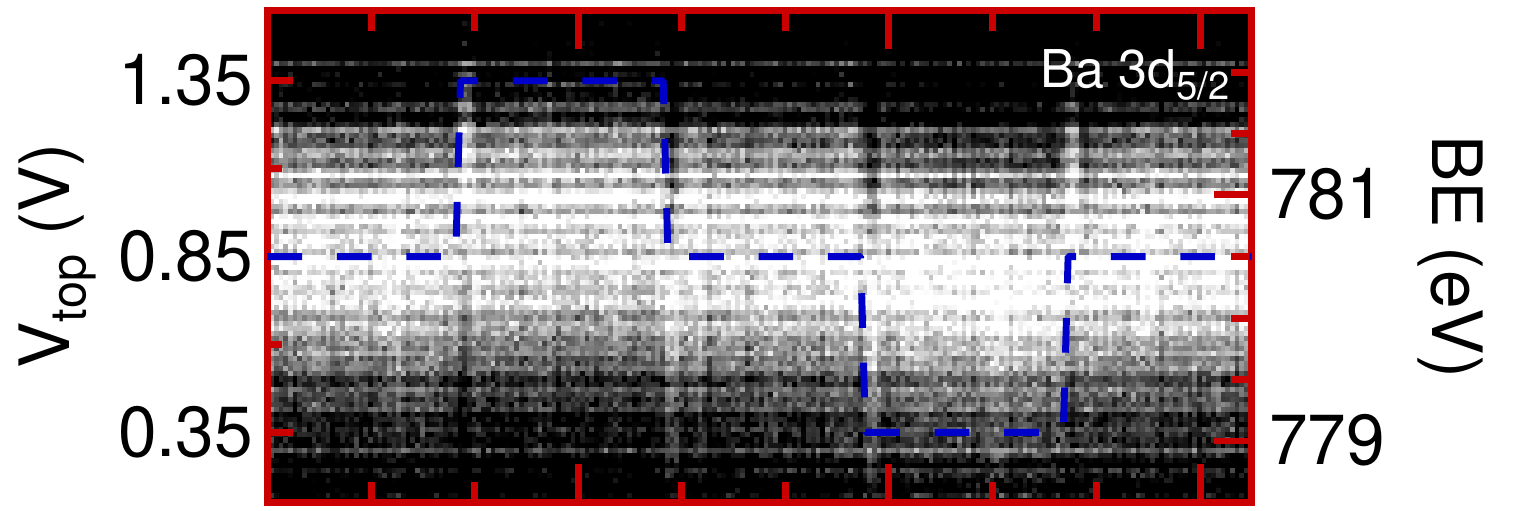}}\\
   	\subfloat{\label{fig:tr_Pt}\includegraphics[scale=0.5]{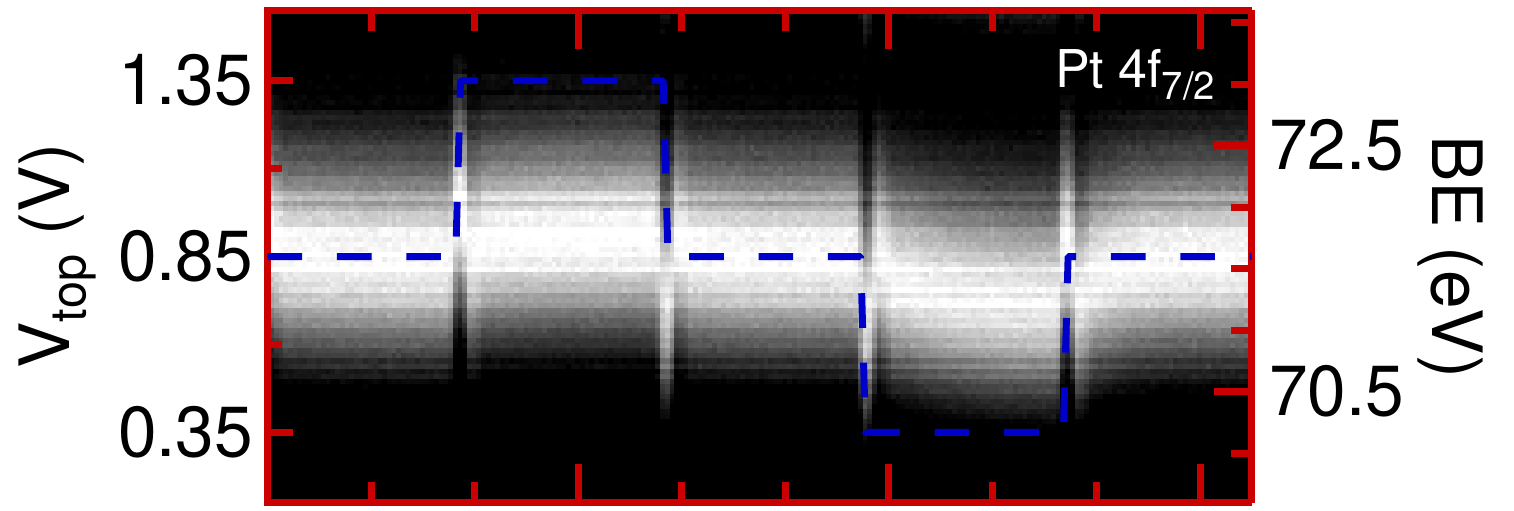}}\\
	\subfloat{\label{fig:tr_Pd}\includegraphics[scale=0.5]{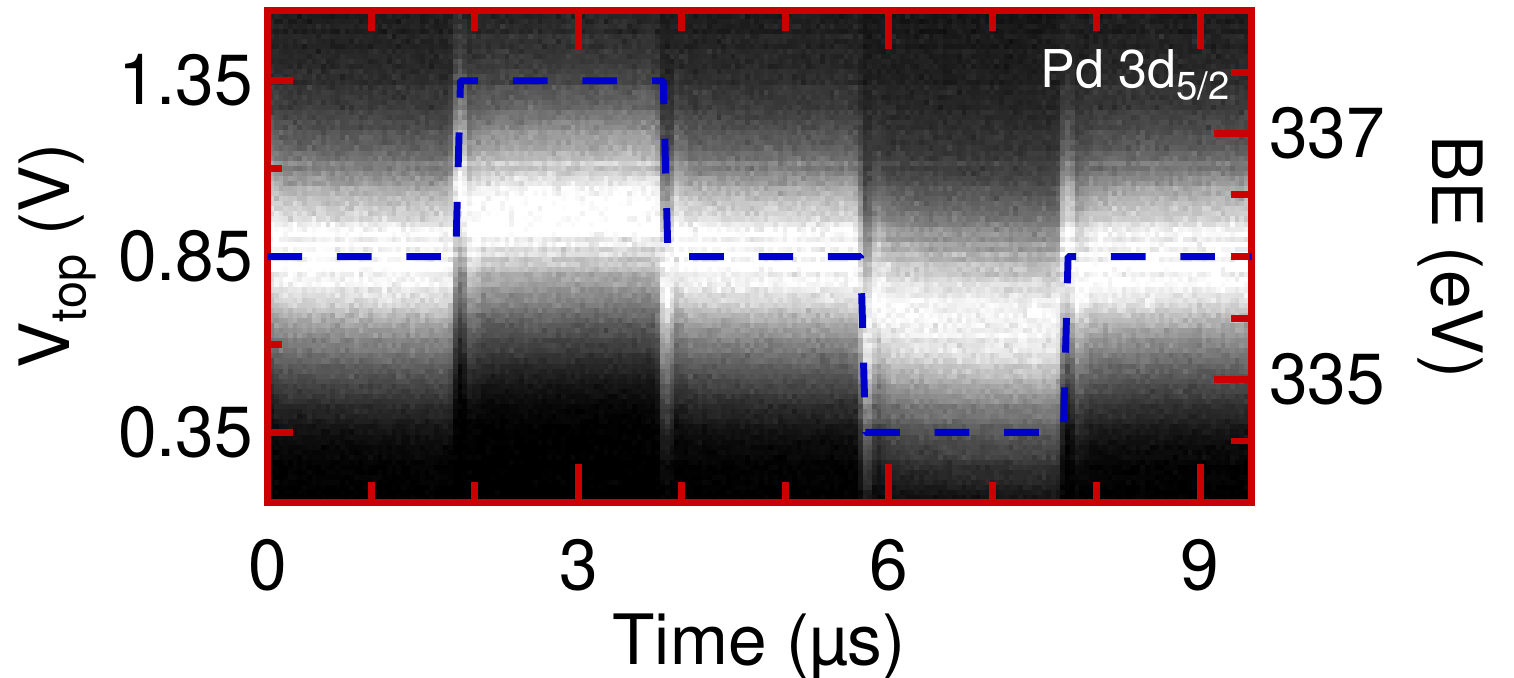}}
  \caption{Photoemitted electron intensity as a function of time and binding energy (BE) for the three core-levels Pd~$3d_{5/2}$, Pt~$4f_{7/2}$ and Ba~$3d_{5/2}$. The blue dotted curve shows the pulse train applied on the top electrode.}
  \label{fig:tr_intensities}
\end{figure}

\begin{figure}[!ht]
  \centering
   	\subfloat{\label{fig:fitting_gaussian_Pt}\includegraphics[scale=0.53]{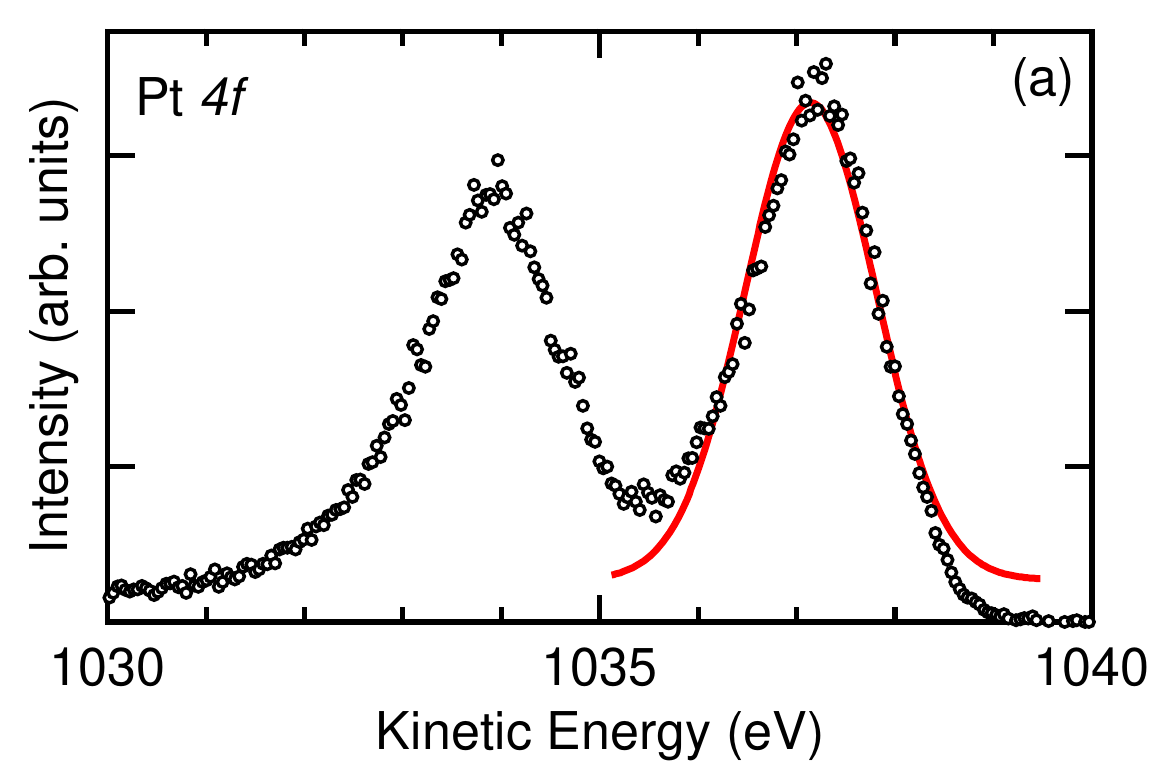}}\\
   	\subfloat{\label{fig:fitting_gaussian_Ba}\includegraphics[scale=0.53]{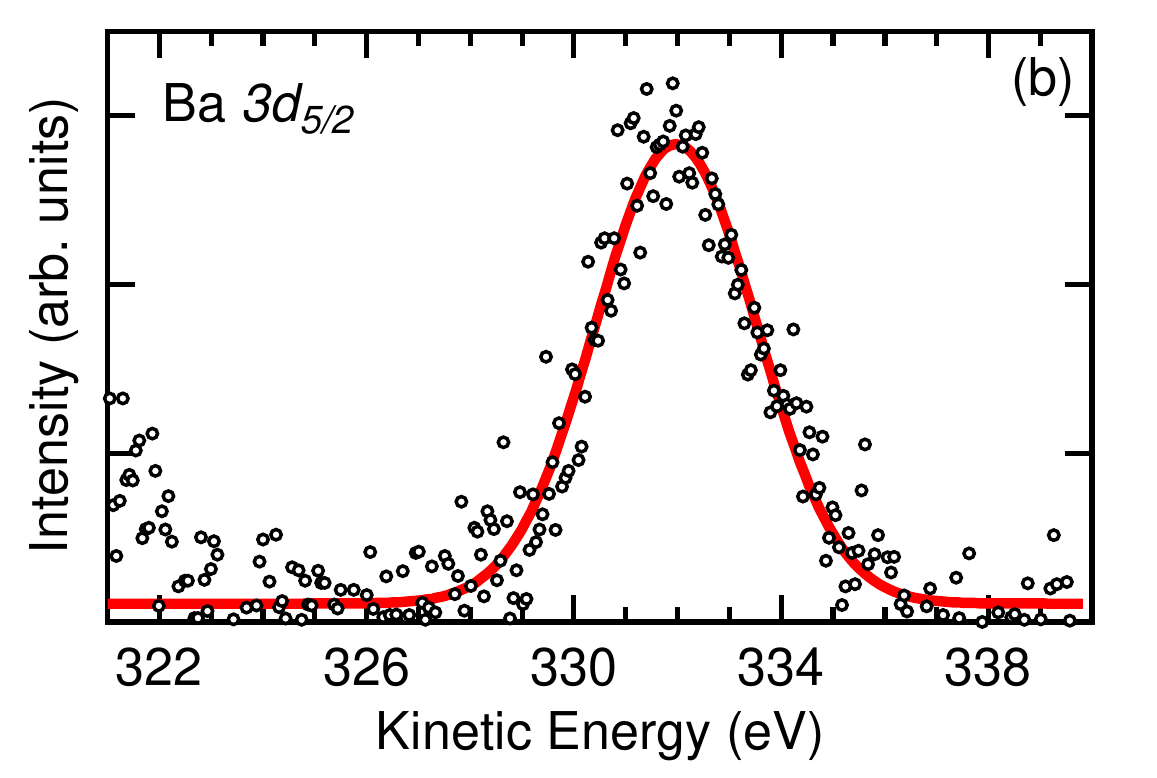}}
  \caption{Example of spectrum of (a) Pt~$4f$ and (b) Ba~$3d_{5/2}$ core-level taken at one time step (black circles) after subtraction of a linear background. The Gaussian fitting lineshape is shown in plain red.}
  \label{fig:fitting_gaussian}
\end{figure}

Figure~\ref{fig:tr_intensities} shows photoelectron intensity as a function of time and binding energy for Pd~$3d_{5/2}$, Pt~$4f_{7/2}$ and Ba~$3d_{5/2}$ core levels corresponding to the pad, electrode and ferroelectric responses to the switching. The shift of the binding energy scale due to the voltage pulses is clearly visible on the core-level spectrum, but this representation is not well-suited for further analysis. We used a fitting procedure which subtracts the secondary electron background and fits the instantaneous spectrum to a Gaussian shape for each time step. It returns the binding energy position of the peak as a function of time and allows a proper comparison of core-level response to the electrical excitation. The width of each peak is kept constant after a preliminary fit procedure using a high-resolution reference spectrum taken with both electrodes grounded. The shape for all core-levels did not change with time except for the spectra close to the rise and fall of the bias pulses. At these moments, the peaks are highly distorted leading to spikes (typical width of 150~ns) in the time-dependent binding-energy curves. A typical spectrum far from the spikes is shown in Figure~\ref{fig:fitting_gaussian} together with the fit. Doniach-Sunjic lineshapes are more suited to metallic peaks~\cite{doniach_many-electron_1970}, however, since we extract binding energy \emph{differences} between identical peaks (in shape) at different time, the use of Gaussian shape is a good approximation and leads to a less time-consuming procedure. We used one component to fit the Ba~$3d$ spectrum though it is known that two components are observed at the Pt/BTO interface~\cite{rault_interface_2013}. However, the signal/noise ratio of the Ba core-level from the buried interface was not sufficient to properly fit two components at every time step (Fig.~\ref{fig:fitting_gaussian_Ba}).

\begin{figure*}[!ht]
  \centering
  	\includegraphics[scale=0.7]{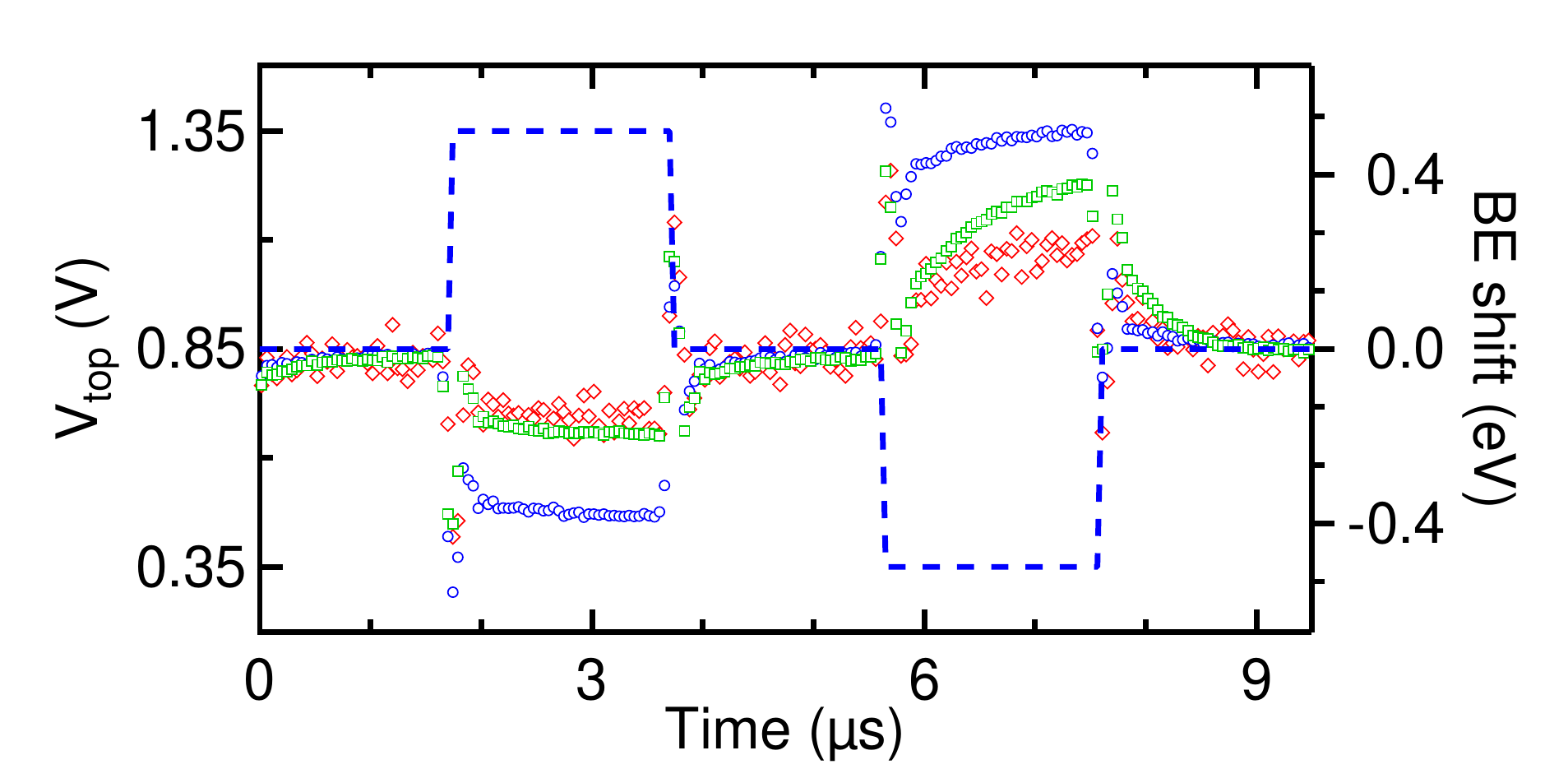}
  \caption{Time-dependent evolution of the binding energy (BE) shifts for the Pd~$3d_{5/2}$ (blue circles), Pt~$4f_{7/2}$ (green squares) and Ba~$3d_{5/2}$ (red diamonds). The blue dotted curve shows the pulse train applied on the top electrode.}
  \label{fig:tr_fittings}
\end{figure*}

Figure~\ref{fig:tr_fittings} shows the time dependence of the binding energies obtained from the fitting procedure. All spectra are referenced relative to their binding energy in the idle state at V$_\mathrm{top} = 0.85$~V. As expected, for the upward (downward) pulse, every core-level shows a decrease (increase) in binding energy due to the more positive (negative) voltage applied on the top electrode in comparison to the idle state. At every pulse rise and fall, spikes are observed in the binding energy. This is likely to be due to a parasitic capacitance with a very small time-constant although we cannot single out its location in the circuit. A capacitance with a time-constant below the rise time of the generator will act as a \textit{differentiator}, \textit{i.e.} will differentiate the signal with respect to time. The time derivative of a step pulse is the delta function we observe as binding energy spikes here. Within the framework of the domain nucleation-domain switching model, Merz distinguished the lead current due to rapid domain nucleation from the current due to domain wall motion and domain switching. Rapid domain nucleation should therefore give rise to a current spike, whereas the response due to domain switching should be broader. Therefore, the core level binding energy spike may also be tentatively correlated with the onset of domain nucleation.

For the non-switching pulse, the core-levels reach a steady state value before the end of the pulse. While Pd follows quite closely the amplitude of the source pulse ($\Delta \mathrm{BE} = 0.4$~eV compared to a pulse $\Delta$V of 0.5~V), Pt and Ba only shift by 0.20~eV. For the P$^{-}$ to P$^{+}$ pulse, the core-level shifts have a higher time constant and each core-level shift is quantitatively different. The $\Delta \mathrm{BE}$ values for the Pd, Pt and Ba core levels after 2~\us are 0.25, 0.37 and 0.50~eV respectively. It is therefore clear that a simple electrostatic model is not sufficient to describe the Pt/BTO interface and that the response of the electronic structure to switches in polarization is non-linear.

\subsection{Simulations}

\begin{figure}[!ht]
  \centering
  	\includegraphics[scale=0.60]{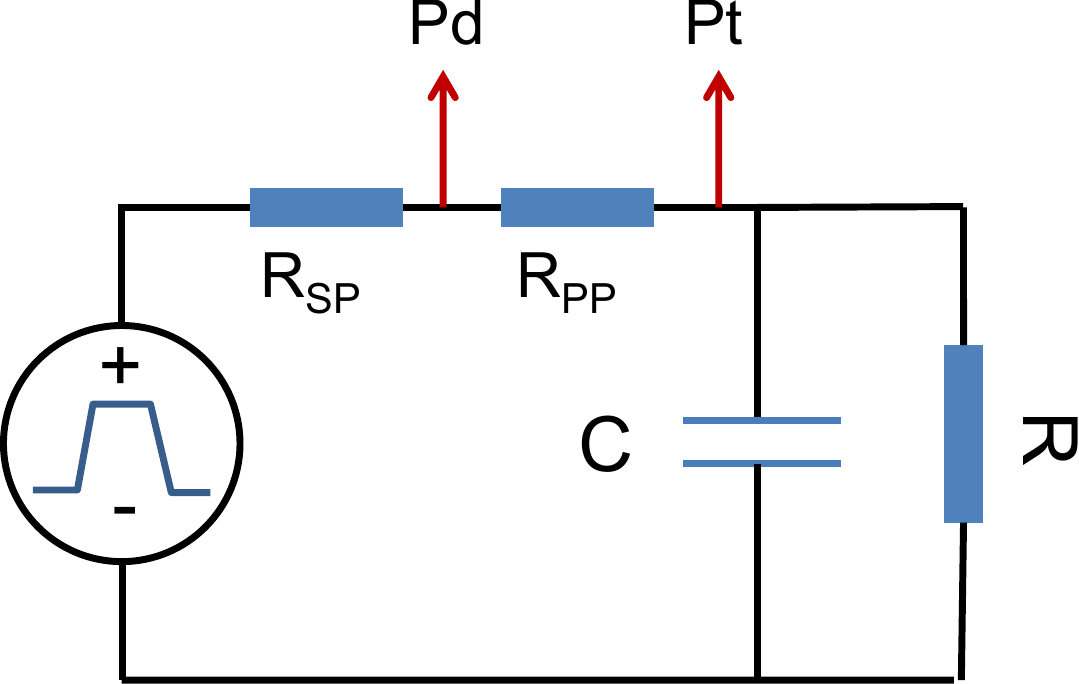}
  \caption{Equivalent circuit used in the PSpice simulation software. The components values are summarized in Table~\ref{tab:tr_values}.}
  \label{fig:tr_circuit}
\end{figure}

\begin{figure}[!ht]
  \centering
  	\includegraphics[scale=0.50]{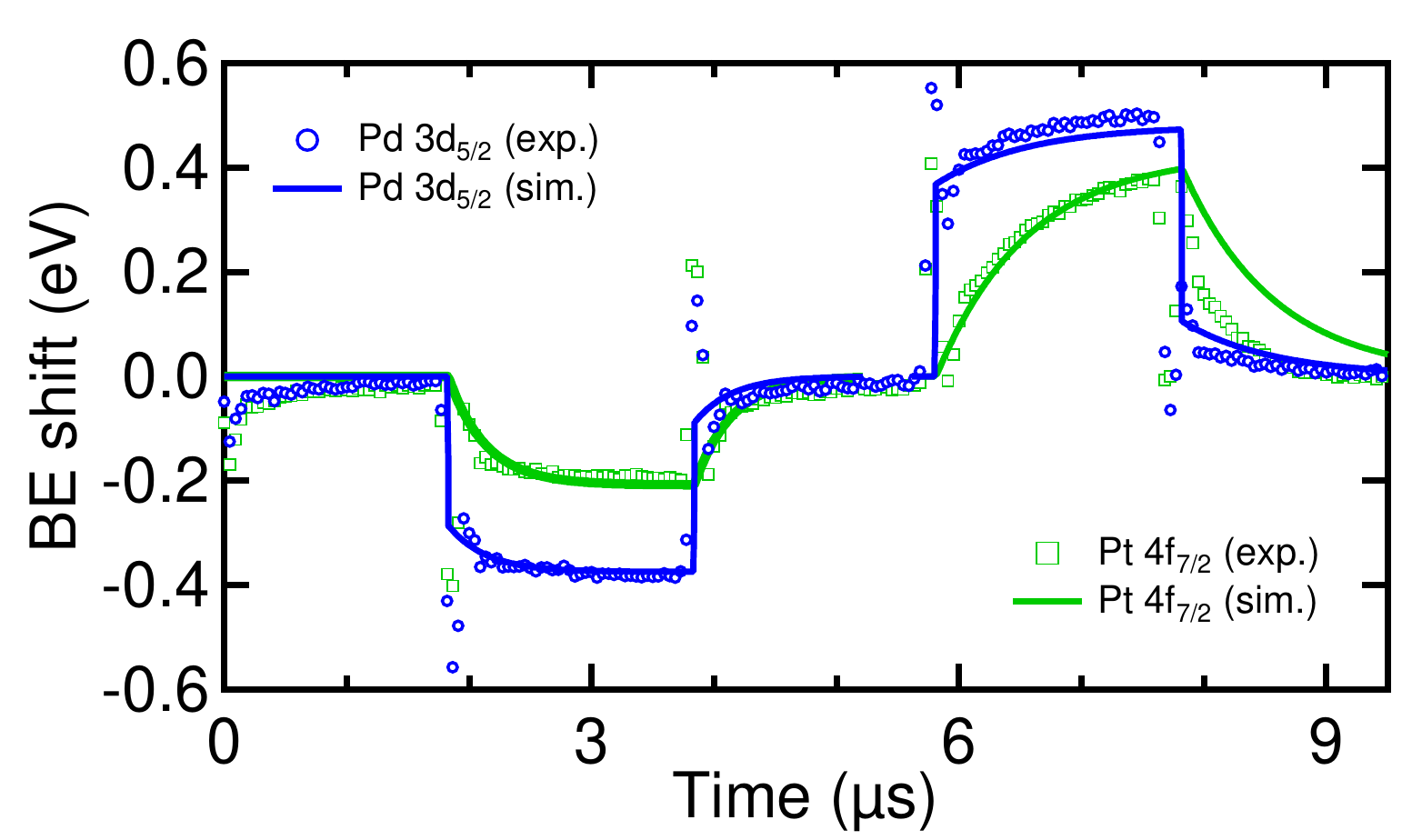}
  \caption{Results of the electrical simulations (plain lines) of the experimental Pd~$3d_{5/2}$ (blue circles) and Pt~$4f_{7/2}$ (green squares) evolution. The simulated curves are generated from the parameters of P$^{-}$ (Table~\ref{tab:tr_values}, right column) from $t = 0$ to $t = 5.8$~\us and from the parameters of P$^{+}$ (Table~\ref{tab:tr_values}, left column) from $t = 5.8$ to $t = 9.2$~\us.}
  \label{fig:tr_simulations}
\end{figure}

We conducted numerical simulations of the time-dependent linear behavior of the capacitor using PSpice software~\footnote{Orcad package from Cadence Design Systems: http://www.cadence.com/products/orcad/pages/default.aspx}. This software can simulate the electrical response of a circuit under a pulse excitation. An upward (downward) pulse identical to the experimental one was used and the circuit response in terms of the potential drop across each component calculated. We assume the resistances between the pulsed source and the palladium pad (source impedance, R\sub{SP}) and between the palladium pad and the platinum electrode (imperfect metal/metal contact, R\sub{PP}) are ohmic. The resistance values should not change with ferroelectric switching and are therefore kept constant. The Pt/BTO/NSTO stack is modeled by a RC circuit of resistance R and capacitance C and of time constant $\tau = RC$. The Pt~$4f_{7/2}$ core-level is therefore a direct probe of the voltage across the capacitor. An RC model accounts for the linear dielectric properties but cannot of course directly model the FE switching, intrinsically non-linear. We have therefore used R and C as fitting parameters that change in response to polarization switch. Figure~\ref{fig:tr_circuit} displays the equivalent circuit used in the simulations. The values which give the best fit to experimental data are given in Table~\ref{tab:tr_values} for each polarization state and the results of the simulation are shown in Figure~\ref{fig:tr_simulations}. 

\begin{figure}[!ht]
  \centering
  	\includegraphics[scale=0.50]{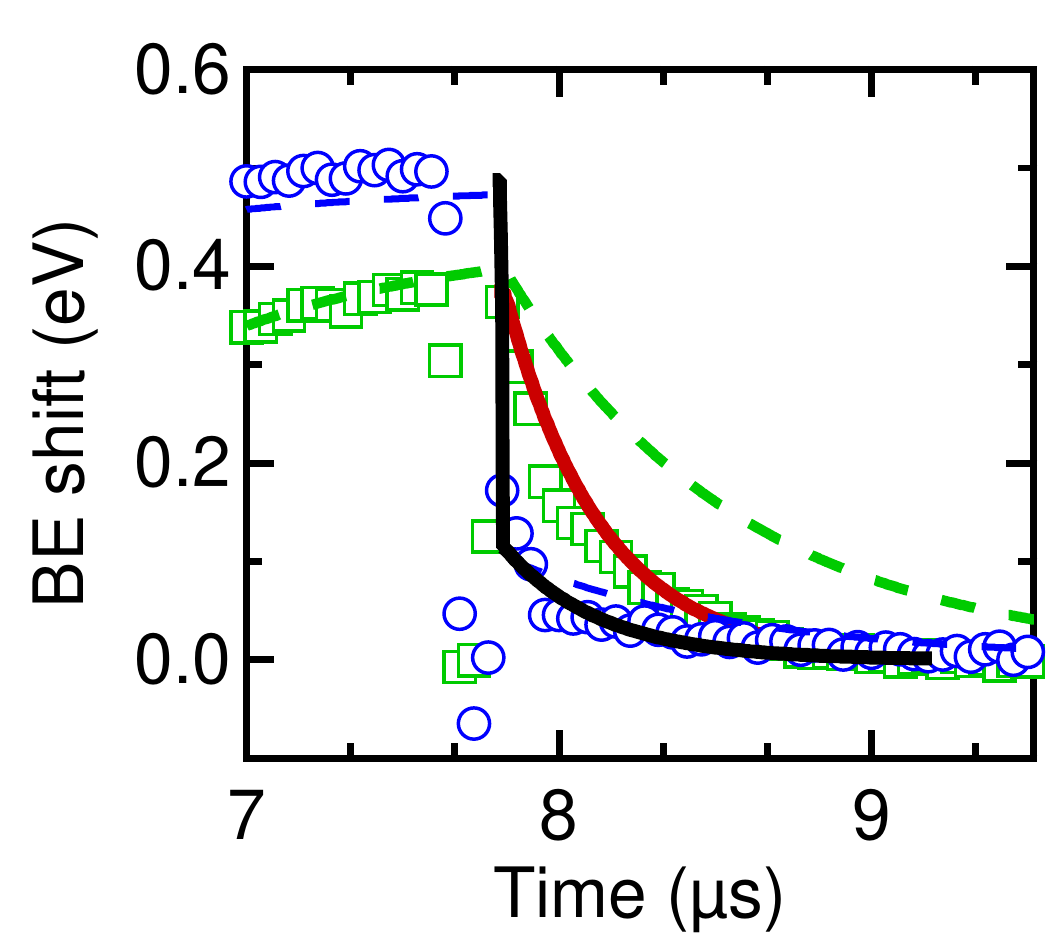}
  \caption{Results of the electrical simulations for the fall sequence of the P$^{-}$ to P$^{+}$ pulse. Pd~$3d_{5/2}$ and Pt~$4f_{7/2}$ experimental binding energy are shown in blue circles and green squares respectively. The simulated plain curves of the fall sequence are generated from the parameters of the P$^{-}$ state (black line for Pd~$3d_{5/2}$, red line for Pt~$4f_{7/2}$). The simulated dotted curves are generated from the parameters of the P$^{+}$ state.}
  \label{fig:tr_simulations_zoom}
\end{figure}

\begin{table}[!ht]
	\centering
	\begin{tabular}{|l r|c|c|}  
	\hline
		& & V\sub{top} = 0.35~V & V\sub{top} = 0.85~V  \\
		& & (P$^{+}$) & (P$^{-}$) \\
	\hline
	R\sub{SP} &($\Omega$)	& 100 	& 210	\\
	\hline
	R\sub{PP} &($\Omega$)	& 280	& 280	\\
	\hline
	R &($\Omega$)	& 2200	& 350	\\
	\hline
	C &(nF)	& 2.3	& 1.5	\\
	\hline
	$\tau$ &(\us) & 5 & 0.5 \\
  	\hline
	\end{tabular}
\caption{Best fit values for the components of the equivalent circuit of Figure~\ref{fig:tr_simulations}, for suffixes refer to text.}
\label{tab:tr_values}
\end{table}
We first discuss the behavior of the Pt and Pd core-levels. It was not possible to properly fit the photoemission data without changing the resistance R\sub{SP} between the palladium pad and the pulse generator. Indeed, when the load (here the Pd/Pt/BTO/NSTO capacitor) on the pulse generator is too small (this is  the case during the non-switching pulse), the source is not able to deliver the full pulse magnitude. In the framework of our simple model, this results in two different values for R\sub{SP}. 

With impedance matching taken into account, the model circuit fits well the experimental values of the non-switching pulse. The steady state low R is consistent with the high-current measured by static electrical experiment~\cite{rault_interface_2013}. For the switching pulse P$^{-}$ to P$^{+}$, the Pt/BTO/NSTO stack resistance R, is higher, consistent with the lower current through the capacitor measured in Ref.~\onlinecite{rault_interface_2013}. The high to low current behavior when switching from P$^{-}$ to P$^{+}$ is due to the increase of the electron barrier height at the bottom interface (BTO/NSTO) of the capacitor (see Figure 7 of Ref.~\onlinecite{rault_interface_2013} for the polarization-dependent band lineups of the Pt/BTO/NSTO heterostructure).

Turning now to the transient behaviour, for the non-switching pulse P$^{-}$ to P$^{-}$, both rise (charge) and fall (discharge) sequence fits well the experimental binding energy shifts. However, for the P$^{-}$ to P$^{+}$ pulse, although the P$^{+}$ simulation curve fits well the rising edge of the core level binding energies, it does not reproduce at all the falling edge because of the time-resolution of our experiment compared with the electron dynamics of polarization switching. At the end of the downward pulse, when the voltage goes from V$_\mathrm{top} = 0.35$~V back to $0.85$~V, the BTO film switches back to the P$^{-}$ state. The dynamic of polarization switching in thin films can be in the nanosecond range depending on the pulse characteristics~\cite{tagantsev_non-kolmogorov-avrami_2002}, the ferroelectric layer properties~\cite{musleh_alrub_thickness_2011} and the electrodes~\cite{kim_polarity-dependent_2011}. This is faster than our time resolution so that, with respect to the spectral acquisition time, the stack instantaneously switches to the P$^{-}$ state at $t = 7.8$~\us. The simulation of the P+ to P- switch in Fig.~\ref{fig:tr_simulations} is calculated from the parameters obtained for the P$^{+}$ state which predict a time constant ten times larger than for the P$^{-}$ state. Figure~\ref{fig:tr_simulations_zoom} shows a close-up of the switch back to P$^{-}$ following the P$^{+}$ pulse. When the parameters of the P$^{-}$ steady state are used the fit is much better, reproducing quantitatively the experimental data (red line for Pt~$4f_{7/2}$, black line for Pd~$3d_{5/2}$ on Fig.~\ref{fig:tr_simulations_zoom}). The time constant of the RC component ($\tau = RC$) estimated from the steady state fit parameters is 0.5~\us (5~\us) for the P$^{-}$ (P$^{+}$) state.

The switching time is therefore sub-\us, in agreement with literature~\cite{gruverman_piezoresponse_2008, larsen_nanosecond_1991, jo_domain_2007}. The much higher value of $\tau$ deduced from the steady P$^{+}$ state may well reflect the RC circuit load rather than a material property.

\begin{figure}[!ht]
  \centering
  	\includegraphics[scale=0.60]{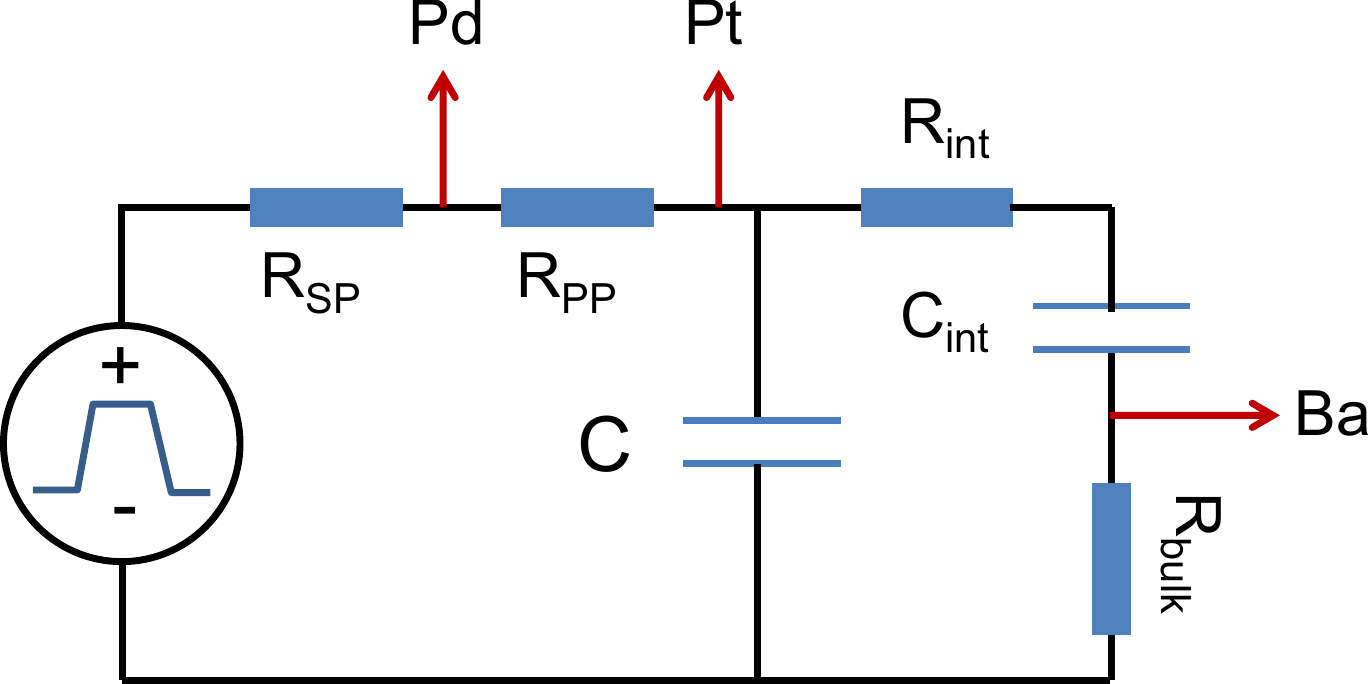}
  \caption{Capacitance model for modeling the barium core-level behavior.}
  \label{fig:tr_circuit_Ba}
\end{figure}

\begin{figure}[!ht]
  \centering
  	\includegraphics[scale=0.50]{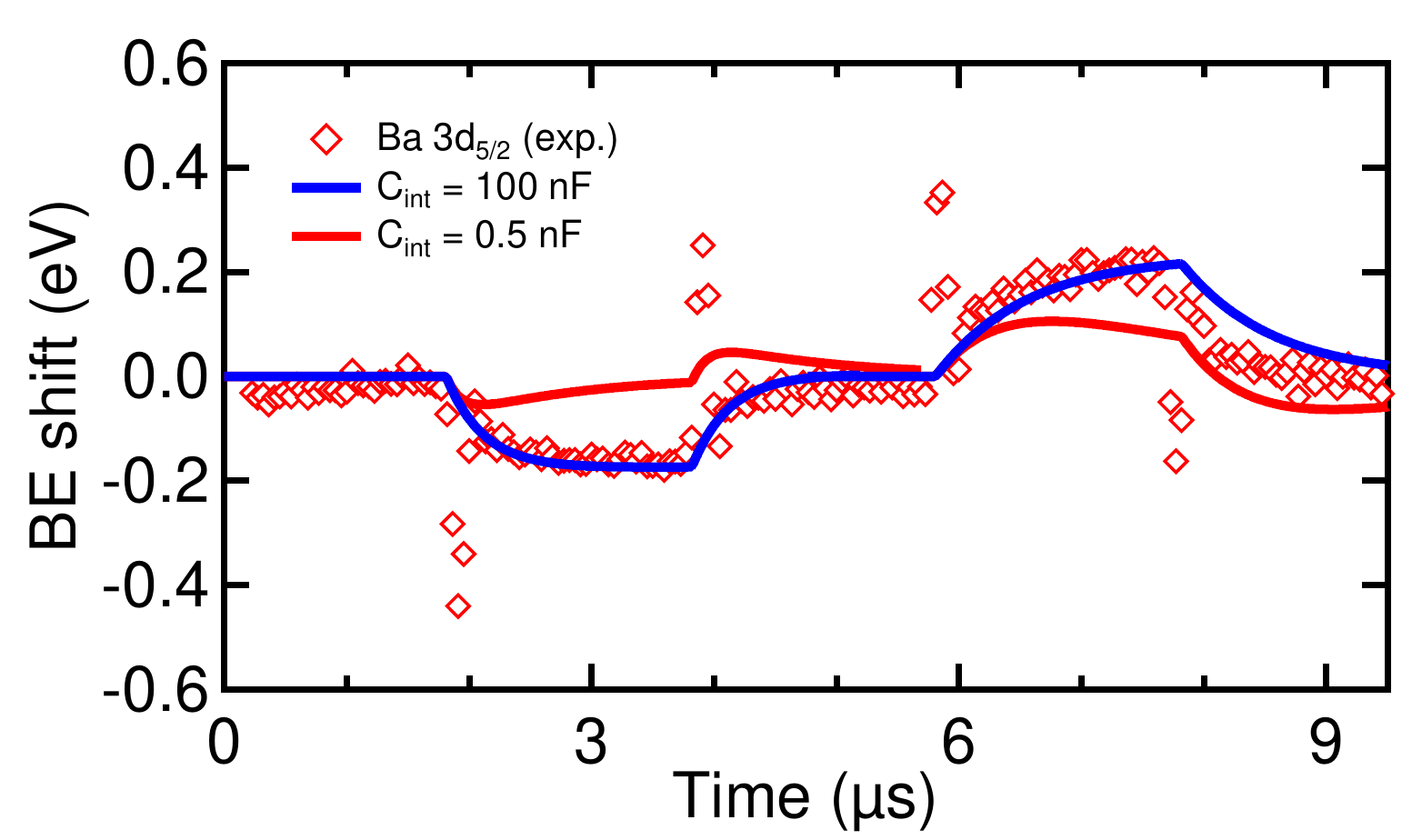}
  \caption{Results of the electrical simulations (plain lines) of the experimental Ba~$3d_{5/2}$ (red diamonds) evolution. The simulated curves are generated with C\sub{int} = 0.5~nF (red line) and C\sub{int} = 100~nF (blue line).}
  \label{fig:tr_simulations_Ba}
\end{figure}

The simulation of the binding-energy response of the barium core-level is more complex because it cannot easily be related to a well-defined probing point in the model circuit. The model must account for the polarization-induced interface dipole but also for the non trivial interface chemistry. As a first approximation, we model the top interface as an additional RC circuit as shown in Figure~\ref{fig:tr_circuit_Ba}. The total capacitor resistance R must be equal to the sum of interface resistance R\sub{int} and remaining bulk resistance R\sub{bulk}. The photoemission probing depth at 1100~eV photon energy is of 3-4~nm so that only electrons emitted from the top interface and the first barium layers are collected. Therefore, a model adding a separate contribution due to the bottom interface is not necessary here. Figure~\ref{fig:tr_simulations_Ba} shows the simulation of the barium binding energy shifts for two extreme values of the interface capacitance C\sub{int}. The model correctly fits the spectroscopic data when the capacitance value is 100~nF, \textit{i.e.} when the interface RC circuit has a time constant well above the pulse train period of 10~\us. For the lowest value (C\sub{int} = 0.5~nF), the simulation is qualitatively incorrect since the simulated signal reaches a maximum before decreasing, which is clearly not the case of the experimental signal. Therefore, to fit the data, the interface capacitance of the model has to be much higher than the bulk capacitance. This is an extremely important result because it places a lower limit on the interface capacitance, C\sub{int}. The Ba~$3d$ transient core level response to the polarization switch therefore confirms the potentially important role of the interface capacitance in thin film ferroelectric based devices.~\cite{stengel_enhancement_2009} We note that this does not necessarily mean that polarization switching time in the whole capacitor stack is 10~\us but that close to the electrode, within the framework of an RC model, there is a significant interface capacitance.

A more sophisticated quantitative analysis is not possible since the RC circuit we use to simulate the stack actually reduces to a simple resistance R\sub{int} at the frequencies used (R\sub{int}$=50~\Omega$ ($1000~\Omega$) and R\sub{bulk}$=300~\Omega$ ($1200~\Omega$) for P$^{-}$ (P$^{+}$) states).

The result is in qualitative agreement with the first-principles-based theoretical results of Stengel \textit{et al.} in Ref.~\onlinecite{stengel_enhancement_2009}. Using a series capacitance model, they found a high value for the interface capacitance relatively to the bulk capacitance (C\sub{int} $\approx 200 \times$C\sub{bulk}) of the Pt/BTO interface. Though the theoretical system (symmetric Pt/BTO/Pt structure) is different from our experimental stack (asymmetric Pt/BTO/NSTO), the semi-quantitative agreement is encouraging and might promote similar experiments with different pulse frequencies or other metal/FE interfaces. The SrRuO\sub{3}/BaTiO\sub{3} interface should show a very different behavior for instance~\cite{stengel_enhancement_2009}. 

Further information on the relationship between the interface electronic structure and capacitance could be obtained by improved signal to noise ratio which would allow a quantitative analysis of the two components of the Ba~$3d$ core level spectrum. It is known that structural changes such as rumpling and interplanar relaxation at the surface can be correlated with the core level shift of the Ba~$3d$ high binding energy component~\cite{pancotti_x-ray_2013}. The use of hard X-ray photoemission would also improve the counting statistics and allow to better identify interface and bulk film contributions to the electronic structure and hence the interface capacitance.

\section{Conclusion}

We have provided a proof of concept for the investigation of realistic devices in operating conditions with time-resolved photoemission spectroscopy. The chemical, electronic and depth sensitivity of core level photoemission is used to probe the transient response of different parts of the upper electrode/ferroelectric interface to voltage pulse induced polarization switching. With the spatial and time resolution available, valuable information on the physics of the interface is accessible. 

The linear dielectric response of the electronic structure is measured using time-resolved photoelectron spectroscopy and agrees quantitatively with a simple RC circuit model. The non-linear response of the electronic structure due to the polarization switch is demonstrated by the time-resolved response of the characteristic core levels of the electrode and the ferroelectric. Adjustment of the RC circuit model allows a first estimation of the Pt/BTO interface capacitance. The experiment shows the interface capacitance is at least 100 times higher than the bulk capacitance of the BTO film, in qualitative agreement with theoretical predictions from the literature. 

Using the temporal structure of x-ray pulses, time-resolved photoemission experiments can reach the picosecond resolution~\cite{bergeard_time-resolved_2011} and thus access valuable information on the electron dynamics of polarization switching itself. Moreover, in our case the spatial resolution is limited by the beam spot size, approximately $100 \times 100$~\ums and prevent from studying realistic microchips. However, the development of photoelectron emission microscopy with \textit{in-situ} bias application, time-resolved detectors and pump/probe setups using high-harmonic generation lasers might lead the way for \textit{in-operando} chemical/electronic analysis of realistic devices at typical operating frequencies.

\begin{acknowledgments}
J.R. is funded by a CEA Ph.D. Grant CFR. This work, partly realized on Nanolyon platform, was supported by ANR project Surf-FER, ANR-10-BLAN-1012 and Minos, ANR-07-BLAN-0312. We acknowledge SOLEIL for provision of synchrotron radiation facilities. We thank J. Leroy, S. Foucquart and C. Chauvet for technical assistance.
\end{acknowledgments}

\bibliography{./biblio_BTO_TR}

\end{document}